# Pseudopotential multiple-relaxation-time lattice Boltzmann simulation of vapor condensation on vertical subcooled walls

*Wandong Zhao[1], Ying Zhang[1*], Ben Xu[2*], Wenqiang Shang[1], Shuisheng Jiang[1]*

1. School of Mechanical and Electrical Engineering, Nanchang University, Nanchang, Jiangxi, 330031, China
2. Department of Mechanical Engineering, University of Texas Rio Grande Valley, Edinburg, TX, 78539, USA

**\*Corresponding authors**
*Tel: +86 (791) 83969634; Email: yzhan@ncu.edu.cn (Ying Zhang)*
*Tel: +1 (956)665-2896; Email: ben.xu@utrgv.edu (Ben Xu)*

**Abstract:** Saturated vapor condensation on the homogenous and heterogeneous subcooled wall under gravity was numerically simulated by the improved pseudopotential phase-change multiple-relaxation-time (MRT) lattice Boltzmann (LB) method. Firstly, the phase-change MRT-LB method coupling finite different method (FDM) to solve the energy equation was verified by the D2 law of the droplet's evaporation, which improve the model could precisely realize the numerical simulation of phase change. Subsequently, effects of wall wettability and subcooling as well as the geometry structure of the heterogeneous on the dynamic behaviors of droplet's growth, coalescence and falling, temperature field and heat flux were investigated via the phase-change model. The results show that for the homogeneous cooler wall, it is easier for dropwise condensation formation with decreasing the wettability of the subcooled surface, and the condensate droplets on the vertical surface are easier to coalesce and fall under the gravity. It was found that in the condensation process, the saturated vapor mainly converges near the triple-phase contact line and leads to the further growth of the condensate droplets, and there is a higher heat flux and much more isotherm folds around the triple-phase contact line than other regions. The subcooled wall with hydrophilic-hydrophobic spacing distribution is more conducive to dropwise condensation; at small subcooling, the transient average heat flux remains almost unchanged, and with the occurrence of condensation, the heat flux rapidly increases, but as wall subcooling increases further, the increases of the heat flux decreases, and when the degree of wall subcooling further increases, the transient heat flux decreases first and then increases. It also can be concluded that increasing the length of the hydrophilic spots on the heterogeneous wall, the dropwise condensation will be easier to form, but too much hydrophilic spots will not contribute to the fall of the droplets.

# 1. Introduce

Condensation process in gas-liquid phase-change is a very complicated and interesting physical phenomenon, which widely appears in natural phenomena (e.g., dew formation, raindrop formation [1, 2]). At the same time, as the most effective heat transfer model, it is also widely applied in industrial engineering and production (e.g., pulsating heat pipe operation process, thermal power generation, the production process, as well as the heat exchange process in refrigeration and air conditioning [2-5]). Therefore, how to improve and enhance the heat exchange efficiency in condensation process and understand the mechanism of condensation has attracted a vast number of researchers [5]. Over the past decades, a lot of scholars have studied the mechanism of condensation by theory and experiment.

Pioneering scientific researcher Nuselt [6] was the first scholar to carry out a large amount of research on the filmwise condensation process, and obtained the distribution equation of temperature and velocity in the laminar condensation process by solving the N-S equation and considering the boundary layer theory. Rose [7] systematically reviewed the research progress of dropwise condensation from experimental and theoretical perspective. They mainly focused on the development of heat transfer tool measurements, the mechanism and theory of condensation, and the transition from dropwise condensation to filmwise condensation, and it was revealed that the drop-by-drop condensation heat transfer coefficient is not constant, but the transition from dropwise condensation to filmwise condensation cannot yet be clearly understood. Vemuri and Kim [8] experimentally and theoretically studied the droplet condensation. They mainly investigated the effect of SAM coating on the condensation characteristics, and the results show that materials with long-chain hydrophobic groups have negligible heat transfer resistance, therefore hydrophobic protective coating is more durable. Leach et al. [9] experimentally observed the kinetics of the droplet growth of vapor condensation and the results showed that the growth of small droplets mainly through the adsorption of vapor in the substrate and spread around the droplets, and they also concluded that the high heat transfer efficiency of the dropwise condensation is because the area of the unit substrate occupied by the small droplets is much larger than the large droplets, therefore the nucleation density can be increased to achieve high heat exchange efficiency. Ma et al. [10] experimentally investigated the condensation process of mixture gas with different percentages of vapor. They focused their attention on the interface effect, and introduced two-phase interface surface free energy during the research process, and the result showed that increasing the wall subcooling will lead to a high surface free energy. Most of the above scholars studied the condensation process via experiment. However, it is very difficult for the experiment to show the basic physics process of temperature field, the velocity vector field and the streamline inside the droplet during the vapor condensation process.

In recent years, due to the development of the theoretical model of the flow and heat transfer and the improvement of the computation fluid dynamics, the utilization of numerical simulation to reveal the phenomenon and mechanism of condensation has been adopted by the majority of scientific researchers.

Li [11] used computational fluid dynamics to numerically simulate the condensation of different percentages of steam in the air in a vertical cylindrical tube, and the results showed that the coolant channel will limit the heat transfer of high quality fractional steam while the average axial velocity will drop rapidly as the water vapor condenses. Da Riva and Del Col [12] discussed film condensation in a mini-channel using the volume of fluid (VOF) method, and the result showed that under the influence of gravity, the film thickness becomes thicker from the inlet to the outlet of the pipe. Lee et al. [13] also employed the VOF method to simulate the downflow condensation in a circular tube, and the results showed that the temperature profile in the condensation process has an irregular shape with a steep gradient.

Most of these scholars realized the numerical simulation of condensate process by directly solving the N-S equation and the energy equation. However, the traditional method is very difficult to deal with the gas-liquid two-phase interface and the evolution of the gas-liquid interface during the condensation process, especially when the evolution of two-phase interface is rapid [14]. At the same time, the discretization of partial differential equations is very complicated, and the demand of computation resource is high [15].

In recent decades, the multiphase pseudopotential lattice Boltzmann (LB) method, due to the free generation of the two phase interface, does not require special treatment of the interface, and the governing equation is a linear equation, which attracts a large number of scholars to applied this method to study the multiphase flow problem [15]. Therefore, some scholars added energy equation based on the pseudopotential LB model, thus resulting in a series of phase-change LB methods. Historically, Zhang and Chen [16] first proposed the gas-liquid phase change LB method based on the multiphase diffusion interface energy equation. Subsequently, Gong and Cheng [17, 18] also established an improved double-distribution single-relaxation-time (SRT) phase change LB model. Liu and Cheng [19] adopted the phase-change model to numerically simulate the film condensation process on the vertical subcooling wall and obtained the contour distribution of temperature and velocity during the condensation process. Further, they also utilized the phase-change model [17, 18] to simulate the dropwise condensation process of saturated steam under the hydrophobic wall surface with different wetting characteristics, and the results showed that increasing the hydrophobic of surface leads to the condensate droplets detached early [20]. Recently, Li and Cheng [14] performed numerically simulate on the saturated steam condensation process in a heterogeneous subcooling horizontal wall under gravity, and investigated the droplets growth, coalescence and departure during condensation based on the Gong and Cheng's model. The velocity field, and temperature field as well as local flux distribution in the process of condensation were also investigated, and the results showed that as wall subcooling decreases, the departure diameter and average cycle time of the condensed droplets will decrease. However, in this physical model, an artificial local hydrophilic point was added to the cooler surface, which induces the condition of condensation droplet formation the location of droplet growth during condensation.

Furthermore, the Gong and Cheng's phase-change model [17] employed Bhatanagar-Gross-Krook (BGK) collision operator [21] to solve the flow and temperature field, therefore, the numerical stability and the selection of the fluid properties of this model are very limited [22]. At the same time, some scholars [23-25] found that the double-distribution phase-change LB model cannot be completely restored to the macroscopic convection diffusion energy equation after multi-scale expansion through Chanpman-Enskog, subsequently, some researchers have developed the LB phase-change model coupled with the finite difference method (FDM) to solve the temperature field accordingly.

Based on the above ideas, Li et al. [26] proposed hybrid thermal LB method, which added a multiple-relaxation-time (MRT) collision operator and FDM to solve the temperature field in the original Shan-Chen [27] pseudopotential multiphase flow model, and the numerical simulation of the gas-liquid phase-change process was realized. They numerically simulated the modal boiling and nucleate boiling on the horizontal wall, and obtained the boiling heat transfer curve via the improved hybrid phase-change LB method. Further, Li successfully simulated the self-driving phenomenon of the droplets dropping on the hot serrated surface [28] and the droplet evaporation on the heterogeneous surface [29] by using the phase–change model, and concluded that the Marangni effect caused by the temperature difference between the droplet is the main reason to promote the movement of the droplets and the dynamic behaviors of "stick-slip-jump" of the droplet evaporation process.

In this paper, an improved thermal pseudopotential MRT-LB model was developed based on the above scholars' research on the phase-change model. It is worth mentioning that this model does not need to add special artificial boundary treatments such as temperature and density fluctuation and structural wall during the phase-change process. At the same time, after reviewing the previous literature, we found that there are few studies on the saturated vapor condensation on the vertical wall under the effect of gravity. Therefore, in this paper, the dynamic behaviors of vapor condensation on the subcooled homogeneous and heterogeneous wall under vertical gravity was numerically simulated using the improved pseudopotential MRT-LB phase change model. Firstly, the MRT-LB phase-change model was deeply introduced, and then the improved model was validated by droplet evaporation and wall wetting characteristics. Subsequently, the phase-change model was employed to numerical investigate the effects of wall wetting characteristics and wall subcooling (Ja number) and hydrophilic-hydrophobic structure parameters of the wall on the dynamic behaviors of droplets' growth, coalescence, falling along the wall and the heat flux during the vapor condensation on the vertical subcooled wall under the gravity

## 2. Numerical model
### 2.1. Pseudopotential MRT-LB model for fluid flow

The pseudopotential LB model proposed by Shan-Chen [27, 30] is widely applied in multiphase flows. However, the collision equation with BGK [21] is utilized in the original model, which makes the model have some defects in numerical stability and numerical accuracy, therefore, Lallemand and Luo [31] proposed MRT operator in the collision process, which makes the stability of the LB model greatly improved and can deal with the flow process under low viscosity. Recently, LI et al. [32] further improved the source term in the MRT collision matrix, which makes the pseudopotential model maintain thermodynamically consistent in applying the Carnahan-Starling (C-S) equation of state (EOS) while satisfying the large density ratio (e.g., 1000). The flow evolution of density distribution function (DF) of MRT collisions with large density external force term is given as follows [33-37]:

$$f_\alpha(\boldsymbol{x}+\boldsymbol{e}_\alpha\delta_t, t+\delta_t) - f_\alpha(\boldsymbol{x},t) = -\overline{\Lambda}_{\alpha i}(f_i - f_i^{eq})\big|_{(x,t)} + \delta_t(S_\alpha - 0.5\overline{\Lambda}_{\alpha\beta}S_\beta)\big|_{(x,t)} \quad (1)$$

where $f$ and $f^{eq}$ represent the particle DF and the equilibrium DF respectively, $\delta x$ and $\delta t$ are the lattice space step and the time step, and both were set to be equal to 1, so $c = \delta x/\delta t = 1$ [26]. $e_a$ denotes the discrete velocity along the direction of $a$. In this study, the D2Q9 model was adopted, so the discrete velocity can be given as [38]:

$$\boldsymbol{e}_i = \begin{cases} (0,0), & i=0 \\ (1,0)c,(0,1)c,(-1,0)c,(0,-1)c & i=1-4 \\ (1,1)c,(-1,1)c,(-1,-1)c,(1,-1)c, & i=5-8 \end{cases} \quad (2)$$

At the same time, $\overline{\Lambda} = \boldsymbol{M}^{-1}\Lambda\boldsymbol{M}$ is the collision matrix, $\boldsymbol{M}$ is the orthogonal transfer matrix, and $\Lambda$ is the diagonal relaxation matrix in the moment space, which can be defined as [32, 39]:

$$\begin{aligned}\Lambda &= diag(s_0, s_1, s_2, s_3, s_4, s_5, s_6, s_7, s_8) \\ &= diag(\tau_\rho^{-1}, \tau_e^{-1}, \tau_\varsigma^{-1}, \tau_j^{-1}, \tau_q^{-1}, \tau_j^{-1}, \tau_q^{-1}, \tau_\upsilon^{-1}, \tau_\upsilon^{-1})\end{aligned} \quad (3)$$

where $S_1 = S_2$, $S_3 = S_5$, $S_7 = S_8$. The flow non-dimensional relaxation time is defined:

$$\tau_\upsilon = \frac{1}{s_7} = \upsilon/c_s^2 + 0.5 \tag{4}$$

where $\upsilon$ is the kinematic viscosity of the fluid.

The viscosity relaxation time in the MRT-LBM is determined by:

$$\tau_v = \tau_g + \frac{\rho - \rho_g}{\rho_l - \rho_g}(\tau_l - \tau_g) \tag{5}$$

where subscripts $g$ and $l$ denote the gas phase and liquid phase, respectively.

By linear transformation, the DF $f$ can be converted to the moment space $m = M \cdot f$, $m^{eq} = M \cdot f^{eq}$. The moment space DF $m$ and the equilibrium DF $m^{eq}$ are determined by [31]:

$$\underbrace{\begin{bmatrix} m_0(\rho) \\ m_1(e) \\ m_2(\varepsilon) \\ m_3(j_x) \\ m_4(q_x) \\ m_5(j_y) \\ m_6(q_y) \\ m_7(p_{xx}) \\ m_8(p_{xy}) \end{bmatrix}}_{m} = \underbrace{\begin{bmatrix} 1 & 1 & 1 & 1 & 1 & 1 & 1 & 1 & 1 \\ -4 & -1 & -1 & -1 & -1 & 2 & 2 & 2 & 2 \\ 4 & -2 & -2 & -2 & -2 & 1 & 1 & 1 & 1 \\ 0 & 1 & 0 & -1 & 0 & 1 & -1 & -1 & 1 \\ 0 & -2 & 0 & 2 & 0 & 1 & -1 & -1 & 1 \\ 0 & 0 & 1 & 0 & -1 & 1 & 1 & -1 & -1 \\ 0 & 0 & -2 & 0 & 2 & 1 & 1 & -1 & -1 \\ 0 & 1 & -1 & 1 & -1 & 0 & 0 & 0 & 0 \\ 0 & 0 & 0 & 0 & 0 & 1 & -1 & 1 & -1 \end{bmatrix}}_{M} \underbrace{\begin{bmatrix} f_0 \\ f_1 \\ f_2 \\ f_3 \\ f_4 \\ f_5 \\ f_6 \\ f_7 \\ f_8 \end{bmatrix}}_{f}, m^{(eq)} = \begin{bmatrix} \rho \\ -2\rho + 3(j_x^2 + j_y^2) \\ \rho - 3(j_x^2 + j_y^2) \\ j_x \\ -j_x \\ j_y \\ -j_y \\ j_x^2 - j_y^2 \\ j_x j_y \end{bmatrix} \tag{6}$$

where $j_x$ and $j_y$ are equal to $\rho u_x$, $\rho u_y$, respectively.

In the MRT-LB model, the collision is calculated in the moment space, while the streaming process is carried out in the velocity space. With the help of the Eq. (3) and Eq. (6), the right-hand side of Eq. (1) can be obtained as [40]:

$$m^* = m - \Lambda(m - m^{eq}) + \delta_t (I - \frac{\Lambda}{2})\overline{S} \tag{7}$$

where $I$ is the unit tensor, $\overline{S} = MS$ is the forcing term in the moment space, and $S = (S_0, S_1, S_2, S_3, S_4, S_5, S_6, S_7, S_8)^T$.

In the LB model, the streaming process can be expressed as:

$$f_i(\mathbf{x} + \mathbf{e}_i \delta_t, t + \delta_t) = f_i^*(\mathbf{x}, t) \tag{8}$$

where $f^* = M^{-1} m^*$. In order to realize the thermodynamic consistency, Li et al. [32] further improved the source term in the Eq. (7), which can be defined in the moment space as:

$$\overline{S} = \begin{bmatrix} 0 \\ 6(u_x F_x + u_y F_y) + \dfrac{12\varpi |F_m|^2}{\psi^2 \delta_t (\tau_e - 0.5)} \\ -6(u_x F_x + u_y F_y) + \dfrac{12\varpi |F_m|^2}{\psi^2 \delta_t (\tau_\varsigma - 0.5)} \\ F_x \\ -F_x \\ F_y \\ -F_y \\ 2(u_x F_x - u_y F_y) \\ (u_x F_y - u_y F_x) \end{bmatrix} \quad (9)$$

where $\varpi$ is applied to keep the numerical stability. And $F_m = (F_{mx}, F_{my})$ is the force acting on the pseudopotential model. For the MRT-LB model, the corresponding macroscopic density and velocity are determined by:

$$\rho = \sum_i f_i, \quad \rho \mathbf{v} = \sum_i e_i f_i + \dfrac{\delta_t F}{2} \quad (10)$$

where $F = (F_x, F_y)$ is the total force, which includes the interaction force between the particles $F_m$, the interaction force between fluid and virtual solid $F_{ads}$ to realize the different wall wetting characteristic and gravitational force $F_g$.

In the pseudopotential model, Shan-Chen [30] employed the interaction particle force, which is the key to mimic the multiphase separation. For the fluid-fluid interaction force in the single-component (SC) multiphase model for fluid flow can be given by:

$$F_m = -G\psi(\mathbf{x},t)\left[\sum_i w(|e_i|^2)\psi(\mathbf{x}+e_i,t)e_i\right] \quad (11)$$

where $G$ is the interaction strength with a positive (negative) sign for a repulsive (attractive) force between particles, and $w(|e_a|^2)$ is the weight factor [39, 40]. For the case of D2Q9 lattice, the weights factors are $w(1) = 1/3$ and $w(2) = 1/12$. and $\psi$ in Eq. (11) can be defined as [41]:

$$\psi = \sqrt{\dfrac{2(P_{EOS} - \rho c_s^2)}{Gc^2}} \quad (12)$$

where $P_{EOS}$ is non-ideal equation of state.

Recently, Li et al. proposed new solid-fluid force based on $\psi$ and can better achieve the different wetting characteristic, and the new interaction force is given by [42]:

$$F_{ads} = -Gw\psi(\mathbf{x},t)\left[\sum_i w(|e_i|^2)\psi(\rho_w)s(\mathbf{x}+e_i)e_i\right] \quad (13)$$

where $s(\mathbf{x}+e_i)$ is a switch function, which is equal to 0 and 1 for solid and fluid, respectively.

The gravitational force $F_g$ is determined by:

$$F_g(\mathbf{x}) = (\rho(\mathbf{x}) - \rho_v)\mathbf{g} \quad (14)$$

where $\mathbf{g} = (0, -g)$. Subsequently, the total force in the Eq. (10) is $F = F_m + F_{ads} + F_g$.

## 2.2. Energy equation for temperature field

The LB model of the gas-liquid phase-change model based on the diffusion interface was first proposed by Zhang and Chen [16], and the energy equation is expressed as follows:

$$\rho \frac{De}{Dt} = -p\nabla \cdot \mathbf{v} + \nabla \cdot (\lambda \nabla T) \tag{15}$$

where $e = CvT$ is internal energy, $Cv$ is the specific heat capacity at the constant volume, $\lambda$ is the thermal conductivity. Further, the entropy's local equilibrium energy equation in the case of neglecting viscous dissipation is expressed as Eq. (16).

$$\rho T \frac{ds}{dt} = \nabla \cdot (\lambda \nabla T) \tag{16}$$

According to the general relation of entropy, the following relations can be obtained

$$ds = \left(\frac{\partial s}{\partial T}\right)_v dT + \left(\frac{\partial s}{\partial v}\right)_T dv \tag{17}$$

According to the Maxwell relationship:

$$\left(\frac{\partial s}{\partial v}\right)_T = \left(\frac{\partial p}{\partial T}\right)_v \tag{18}$$

Further, based on the chain relation and the definition of specific heat capacity, the following relations can be obtained:

$$\left(\frac{\partial s}{\partial T}\right)_v = \frac{(\partial u / \partial T)_v}{(\partial u / \partial s)_v} = \frac{Cv}{T} \tag{19}$$

with the help of Eqs. (18) and (19), the following Eq. (20) can be obtained from Eq. (17)

$$ds = \frac{Cv}{T} dT + \left(\frac{\partial p}{\partial t}\right)_v dv = \frac{Cv}{T} dT + \left(\frac{\partial p}{\partial t}\right)_v d\left(\frac{1}{\rho}\right) = \frac{Cv}{T} dT - \frac{1}{\rho^2}\left(\frac{\partial p}{\partial t}\right)_v d\rho \tag{20}$$

Thus, the Eq. (16) can be rewritten as:

$$\rho Cv \frac{dT}{dt} = \frac{T}{\rho}\left(\frac{\partial p}{\partial t}\right)_\rho \frac{d\rho}{dt} + \nabla \cdot (\lambda \nabla T) \tag{21}$$

Using the material derivative $D(\cdot)/Dt = \partial_t(\cdot) + \mathbf{v} \cdot \nabla(\cdot)$, Eq. (21) can be converted to the following equation:

$$\frac{\partial T}{\partial t} + \mathbf{v} \cdot \nabla T = \frac{1}{\rho Cv}\nabla \cdot (\lambda \nabla T) + \frac{T}{\rho^2 Cv}\left(\frac{\partial p}{\partial t}\right)_\rho \frac{d\rho}{dt} \tag{22}$$

Further, by using the continuity equation, Eq. (22) can be further rewritten as follows:

$$\frac{\partial T}{\partial t} + \mathbf{v} \cdot \nabla T = \frac{1}{\rho Cv}\nabla \cdot (\lambda \nabla T) - \frac{T}{\rho Cv}\left(\frac{\partial p}{\partial t}\right)_\rho \nabla \cdot \mathbf{v} \tag{23}$$

The right side of the Eq. (23) is marked by $K(T)$. With the help of the second-order Runge-Kutta scheme [43], The time discretization of the temperature field was solved by the following equation:

$$T^{t+\delta t} = T^t + \frac{\delta t}{2}(h_1 + h_2) \tag{24}$$

where $h_1$ and $h_2$ can be determined as:

$$h_1 = K(T^t), h_2 = K(T^t + \frac{\delta t}{2} h_1) \tag{25}$$

In general, the flow process solved by MRT-LB model and temperature field solved by FDM can be coupled by the $P_{EOS}$ in the Eq. (12). Moreover, the parameter $\rho$ and $v$ are determined by the pseudopotential MRT-LB model.

## 3. Model verification

In this part, we will perform numerical verification on the wall wetting characteristic on solid surface and the classic D2 law of saturated droplet evaporation to prove the numerical stability and accuracy of the improved pseudopotential thermal MRT-LB model, because there will be appeared phase-change and triple-phase contact line during the vapor condensation in the neat section. The EOS defined in Eq. (12) is the key to mimic numerical gas-liquid phase-change, and Peng and Laura [41] proposed the EOS for the saturated vapor phase-change considering the diffusion interface. The Peng-Robinson (P-R) EOS is described as follows:

$$P_{EOS} = \frac{\rho RT}{1-b\rho} - \frac{a\varphi(T)\rho^2}{1+2b\rho-b^2\rho^2} \tag{26}$$

where $\varphi(T) = [1+(0.37464+1.54226\omega-0.26992\omega^2)(1-\sqrt{T/T_{cr}})]^2$, and $w$ is equal to 0.344, $a = 0.45724R^2T_{cr}^2/P_{cr}$, and $b = 0.0778RT_{cr}/P_{cr}$. The parameter $T_{cr}$, $P_{cr}$ are the critical temperature and pressure, respectively. Following the Li's work [26], the parameters $a$, $b$ and $R$ are equal to 2/49, 2/21 and 1 respectively, which means the critical temperature $T_{cr}$ is equal to 0.1094. In addition, according to Ref. [26], when $T = 0.86T_{cr}$, the densities of the liquid phase and the gas phase are $\rho_L = 6.5$, $\rho_V = 0.38$ respectively. Unless otherwise specified, the density of the liquid phase and gas phase keep the same in next paragraphs.

### 3.1. Validation of contact angle

The initial droplet was firstly placed on the bottom wall of a lattice calculation domain $Nx \times Ny = 300$ l.u. × 150 l.u. (l.u. represents lattice units), and the contact angle was tested after droplet stabilization. The initial coordinates of the droplet was set as ($Nx/2$, 1), and the radius of the droplet was taken as 50 (l.u.). It is worth noting that the temperature in the computational domain is consistent and set to saturation temperature $T_{sat} = 0.86 T_{cr}$. For the flow boundary conditions, periodic boundary was adopted in the left and right computational domain, and the top and bottom of the computational domain were set to be solid wall, therefore, the half-way bounce back boundary condition was employed [44]. In addition, the fluid-solid interaction force of the bottom wall was calculated by Eq. (13). In order to ensure the stability and convergence of this model, following the suggestions in the Refs. [32, 45], the relaxation times in $\Lambda$ were chosen as follows: $\tau_\rho = \tau_j = 1.0$, $\tau_e^{-1} = \tau_\varsigma^{-1} = 1.1$, $\tau_q^{-1} = 1.1$, and $\tau_\upsilon = 0.8$, furthermore, the triple-phase contact line was calculated by the following two equations:

$$R = \frac{4H^2+L^2}{8H} \tag{27}$$

where $H$ is the maximum height of the droplet, $L$ is the wetting length of the droplet along the surface, and the gas-liquid interface position was defined as $(\rho_l + \rho_v)/2$. Finally, the contact angle of the droplet can be obtained by arc tangent function with the following equation:

$$Tan\theta = \frac{L1}{2(R-H)} \tag{28}$$

It is worth mentioning that when the wall is a hydrophilic surface, $L1$ is equal to $L$, on the contrary, $L1$ is the maximum interface length in the horizontal direction.

Fig. 1 presents the two-phase distribution contour after droplet is stable. It can be seen from the figure that the wall wetting characteristics gradually change from hydrophilic to hydrophobic. Therefore, Eq. (14) can well achieve different wetting characteristics of the surface. At the same time, the relationship between the contact angle and the $Gw$ after the droplet is stable as shown in Fig. 2. It can be seen from the fitting data that the contact angle and the adjustment parameters $Gw$ can satisfy the linear relationship well. In summary, the wall wetting characteristics can be precisely achieved by Eq. (14).

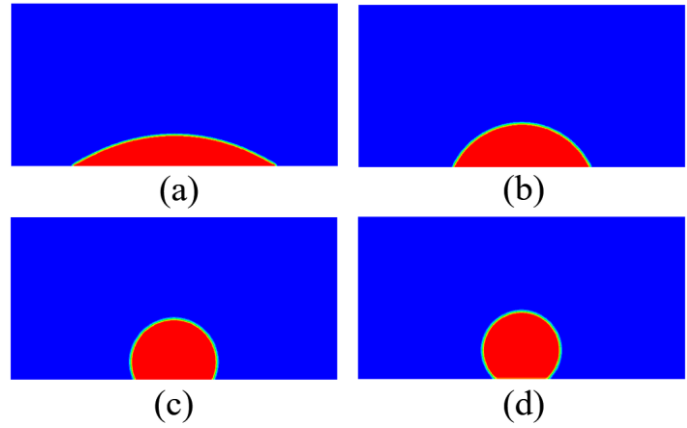

Fig. 1. The steady state of droplet under different wettability walls (Case a $Gw$=-0.2, $\theta$=36.86, Case b $Gw$=-0.2, $\theta$=79.01, Case c $Gw$=-0.2, $\theta$=112.84, Case d $Gw$=-0.2, $\theta$=136.78)

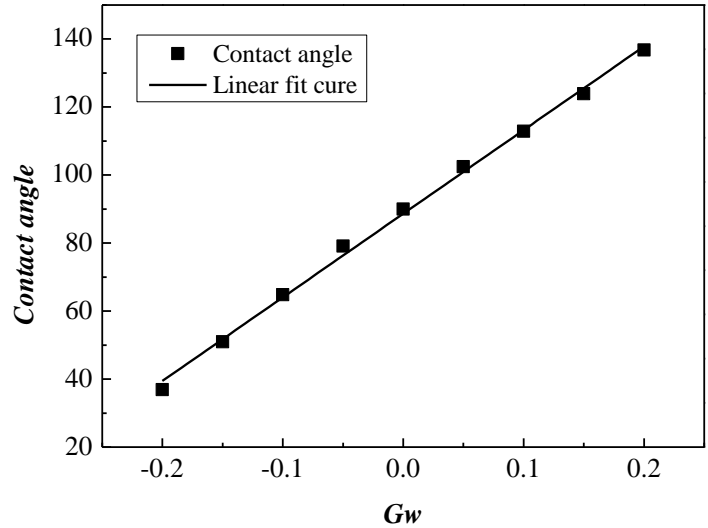

Fig. 2. Relationship between the contact angle and the parameter of $Gw$

### 3.2. Validation of the D2 law

Next, in order to verify the accuracy and reliability of the phase change model in the numerical calculation process, we considered verifying the law of droplet evaporation, namely D2 Law [28, 46, 47]. The initial droplet was placed in the middle of the square domain $Nx \times Ny$ =200 l.u.× 200 l.u., and the droplet initial diameter was set to be $D$ =60 (l.u.), at the same time, the temperature of droplet and the surrounding saturated vapor were set to saturation temperature $T_{sat}$, but the ambient boundary temperature was take as $Tg$, so the droplet slowly evaporates depending due to the temperature gradient. According to the conditions set in D2 law, the physical properties of the fluid remain unchanged during the evaporation process, so we considered the droplet evaporation process with two different thermal conductivities which were equal to Case e: $\lambda = 0.3$ and Case f: $\lambda = 0.6$, and the gas and liquid Kinematic viscosity were both set to be 0.1, and keep $T_{sat}$ =0.86$T_{cr}$, $Tg = T_{cr}$.

Fig. 3 illustrates the droplet evaporation process with two different thermal conductivities. From the figure we can observe that the droplet diameter gradually decreases with the heat transfer. At the same time, we also found that increasing the thermal diffusivity of the fluid leads to a fast evaporation rate, resulting in faster droplet radius reduction. Fig.4 presents the relationship between the droplet diameter and the time step in the two thermal conductivities, where $D1$ represents the transient diameter of the droplet. One can observe from the figure that the square of the ratio of the transient droplet diameter to the initial diameter is linear with the time step, which is in perfect agreement with the D2 Law [28, 46, 47]. Therefore, it can be concluded that the pseudopotential thermal MRT-LM model can well realize the thermal numerical simulation of the two-phase phase change.

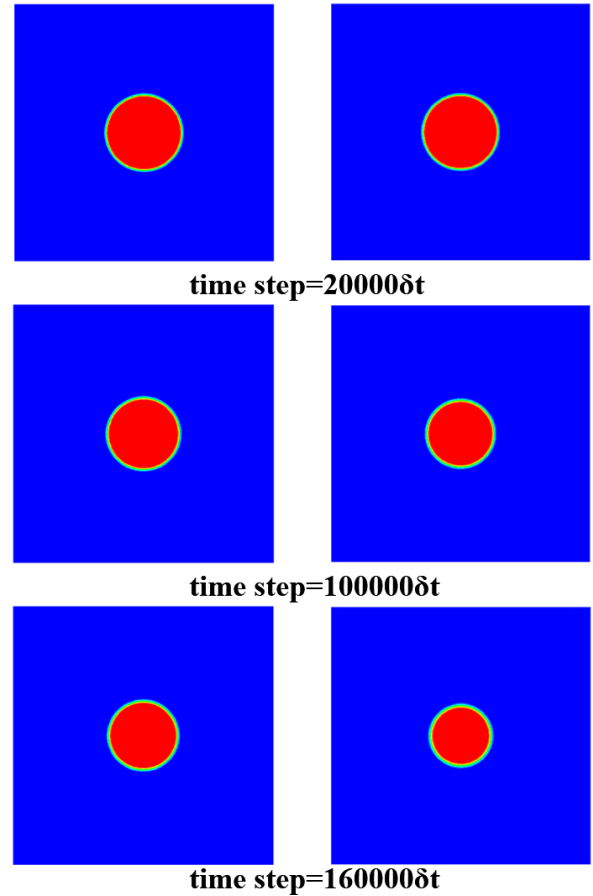

Fig. 3. Droplet evaporation process with different thermal diffusivities, Case e：$\lambda$ =0.3（the left column）,

Case f：$\lambda$ =0.6（the right column）

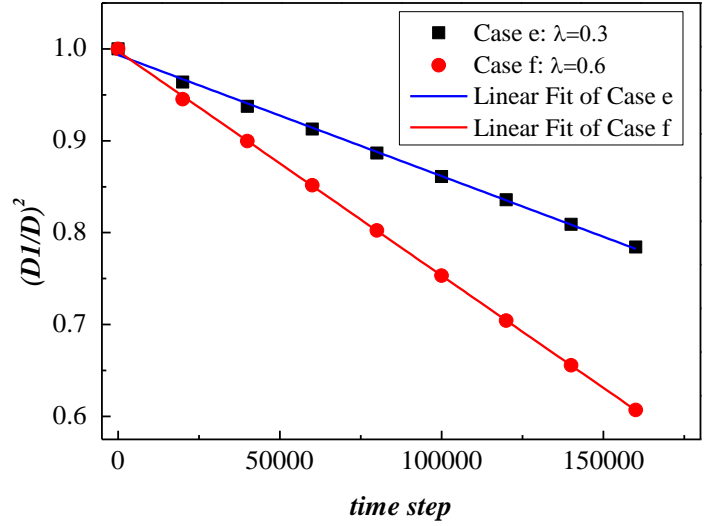

*Fig. 4. Relationship between the transient droplet's diameter and the time step*

## 4. Result and discuss

In this section, the improved pseudopotential thermal MRT-LB model discussed in the previous section was applied to investigate the saturated vapor condensation on the vertical wall. The physical model of saturated vapor condensation, droplets' growth, coalescence and falling in the condensation process under gravity will be discussed in following sections.

*4.1. Computational domain and boundary conditions*

The physical model of saturated steam condensation is shown in Fig. 5. Considering the computational resources and resolution, the grid calculation domain $Nx \times Ny$ =300 l.u. × 900 l.u. was employed to simulate the saturated steam condensation numerically, which means $\lambda_d$ = 300 (l.u.). As shown in Fig. 5, the geometric model was divided into the homogeneous wall of Model A (maintain the same wetting property on the subcooled wall) and the heterogeneous wall surface of Model B (inconsistent wetting characteristics on the subcooled wall). From the Fig.5, the top and bottom walls (marked by red) were adiabatic walls, while the middle area (with blue marks and yellow marks) was set to cooler wall with low temperature ( $T = T_w < T_{sat}$ ). For Model B, the blue area indicates the hydrophobic surface, while the yellow area represents the hydrophilic wall. The computational domain is saturated with dry saturated steam at the initial moment, and the vertical direction considered the gravitational acceleration (0, $g$ ) in the opposite direction of the y axis. In this simulation, the half-way bounce back boundary condition [44] for density DF was employed on the solid wall of the y-direction of the computational domain, while the periodic boundary conditions were imposed on the top and bottom of computational domain.

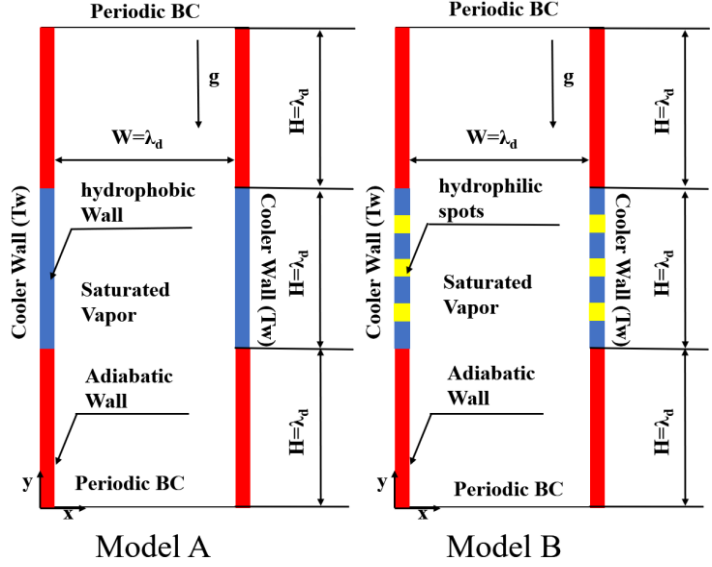

Fig. 5. Schematic of vapor condensation physical model and computation domain

It is worth noting that during the numerical simulation of condensation, the degree of wall subcooling was nondimensionalized by Eq. (29), which is determined by Jacob number ($Ja$) [14].

$$Ja = \frac{Cv(T_{sat} - T_w)}{h_{fg}} \tag{29}$$

where $h_{fg}$ is latent heat, following the Gong and Cheng's model [17, 18], the latent heat of $h_{fg}$ is determined by the following equation:

$$h = [aT \cdot \eta(\omega) \cdot \sqrt{\varphi(T)} \cdot \frac{1}{\sqrt{T \cdot T_{cr}}} + a\varphi(T)] \cdot \frac{1}{2\sqrt{2}b} \ln\left|\frac{2b^2\rho - 2b - 2\sqrt{2}b}{2b^2\rho - 2b + 2\sqrt{2}b}\right| \tag{30}$$

where $\eta(\omega) = (0.37464 + 1.54226\omega - 0.26992\omega^2)$, therefore, the specific latent heat can be calculated by $h_{fg} = h_v - h_l$, and the accuracy of the above equation had been verified by water in the Gong and Cheng's work [17]. In addition, in this simulation, following the Ref. [14], the characteristic length was defined as following, which indicates the ratios of surface tension and buoyancy:

$$l_0 = \sqrt{\frac{\sigma}{g(\rho_l - \rho_v)}} \tag{31}$$

where $\sigma$ is the surface tension of the droplet, which can be calculated by Laplace's law verification, subsequently, the characteristic time can be obtained by Eq. (32).

$$t_0 = \sqrt{l_0 / g} \tag{32}$$

It is worth mentioning that in this simulation, all the physical quantities were calculated by the lattice units, and assumed $\delta x = \delta t = 1$, and the conversion of lattice units and physical units was referred to Ref.[48]. Following the Ref. [14], the non-dimensional parameters between the lattice units and the macroscopic units can be obtained via the following equation:

$$T^* = \frac{T^{lu}}{T_{cr}^{lu}} = \frac{T^{real}}{T_{cr}^{real}}, \quad l^* = \frac{l^{lu}}{l_0^{lu}} = \frac{l^{real}}{l_0^{real}}, \quad t^* = \frac{t^{lu}}{t_0^{lu}} = \frac{t^{real}}{t_0^{real}} \tag{33}$$

where $T_{cr}^{lu}$ =0.1094 is calculated by Eq. (26), and $t_0^{lu}$ =599.33 is determined by Eq. (31) and Eq. (32). Table 1 lists values of different Cases employed for the numerical simulation on a vapor condensation on the vertical subcooled wall.

Table 1. Parameters of different Cases adopted in this paper with computational domain( $Nx \times Ny$ =300× 900)

| Case | $\theta_s$ | $\theta_w$ | Ja | $L_s^* = L_s / l_0$ | $L_w^* = L_w / l_0$ | Number of hydrophilic spots |
|---|---|---|---|---|---|---|
| 1 | - | 115° | 0.244 | - | - | - |
| 2 | - | 135° | 0.244 | - | - | - |
| 3 | - | 155° | 0.244 | - | - | - |
| 4 | 65° | 135° | 0.244 | 0.33 | 3.90 | 3 |
| 5 | 65° | 115° | 0.05-0.26 | 0.33 | 3.90 | 3 |
| 6 | 65° | 135° | 0.05-0.26 | 0.33 | 3.90 | 3 |
| 7 | 65° | 135° | 0.05-0.26 | 0.66 | 3.67 | 3 |
| 8 | 65° | 135° | 0.05-0.26 | 0.99 | 3.40 | 3 |
| 9 | 65° | 135° | 0.05-0.26 | 0.33 | 2.51 | 5 |

*4.2. Growth, coalescence and subsequent fall and distribution of velocity and temperature field of droplets during saturated vapor condensation on homogeneous hydrophobic walls under the gravity*

First, we numerically simulate the dynamic behaviors of the saturated steam condensation on a homogenized subcooling wall. The influence of the wetting characteristics of the hydrophobic surface was mainly analyzed. It should be noted that the implementation of the condensation process on the homogeneous wall has not been affected by any other artificial factors, such as setting the growth core of condensation, temperature fluctuation, and the influence of the wall structure. Therefore, the initial droplet growth in the numerical simulation of condensation is automatically generated on vertical cooler wall.

*4.2.1. Vapor condensation process on vertical walls with different wetting characteristics*

Fig.6 presents the dynamic behaviors of droplets' growth, coalescence and subsequent falling under the homogeneous hydrophobic subcooled wall at a large wall subcooling $Ja$ =0.244 with the contact angle of $\theta_w$ =115°. The red and blue colors marked by the figure represent condensing droplets and saturated vapor, respectively. One can observe from the figure that when t*=6.67, a thin liquid film has formed on the cooler wall due to the greater degree of wall subcooling, and two separate small droplets form at the ends of top and bottom cold region respectively. At the next moment (t*=10.01), the liquid film becomes thicker and the middle liquid film converges to the middle region due to the effect of the surface tension, at the same time there is further smaller droplets growth on the middle subcooled wall as presented by Fig.6(c). When t * = 16.69 and t * = 20.02, further convergence of the liquid film in the middle region can be observed, and the surrounding droplets around liquid film gradually grow in volume and coalesce the little condensation droplets, resulting in the volume of large droplets increase. Finally, under the effects of gravity and surface tension, the droplets begin to fall along the cooler wall, and the contact angle hysteresis formed under the influence of gravity could be clearly observed.

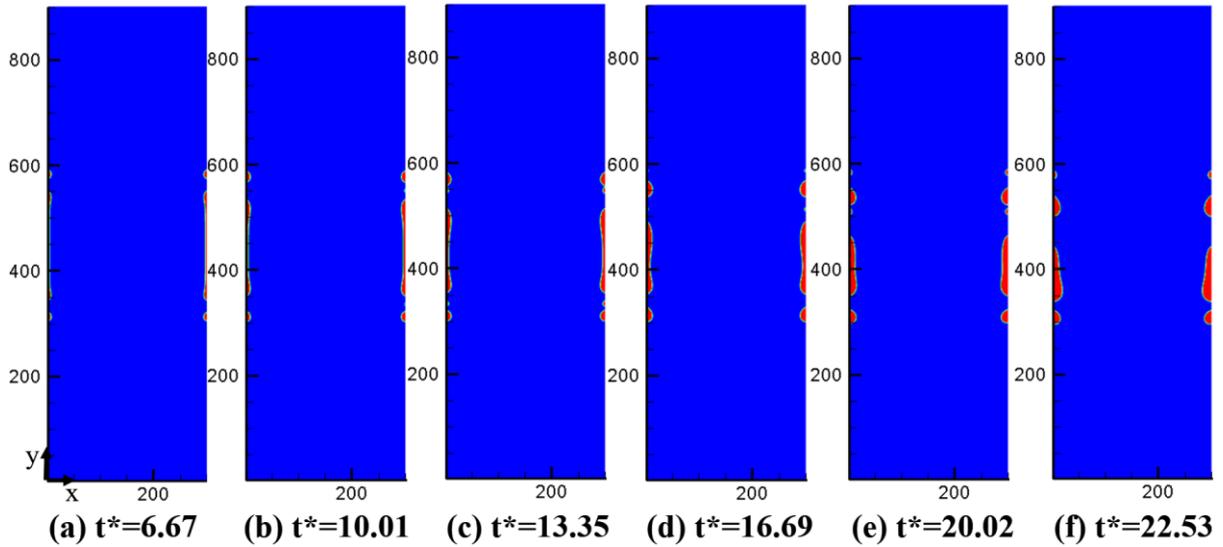

Fig. 6. Snapshots of droplet's growth, coalescence and falling under the homogeneous hydrophobic subcooling wall under gravity(Case #1  Ja =0.244,  $\theta_w$ =115°)

Fig.7 illustrates the dynamic behaviors of condensate droplet's growth and coalescence under the subcooled hydrophobic wall with the contact angle of $\theta_w$ =135° and Ja =0.244. From the figure we can observe that when t* = 6.67, it is similar to the simulation results in the previous Case #1, but there are much tinier droplets generated at the ends of cooler areas. Further at t*=10.01, it can be observed that the rate of condensation of the liquid film formed in the middle region is faster than that of Case #1. This is due to the increase of wetting characteristics of the subcooled surface resulting in lower wall adhesion to the droplet, and the liquid film shrinks faster under the effect of surface tension. At the same time, one can observe from Fig.7 (c) that, more fine droplets are generated around the large droplets. Under the influence of gravity, as indicated by Fig. 7(c), the large droplet at Y=500 merges with the small droplet, resulting in a further increase in the volume of the droplet at the next moment t* = 15.85. Due to the increase of the volume of condensate droplets, the droplets began to fall in the Y direction under gravity, and from Fig.7(e), one can observe that the contact angle hysteresis is clearly formed at the Y=400 droplets, while the smaller droplets still stay in the initial position. The droplet at Y = 350 shown in Fig. 7(f) further coalesces with the downstream droplets and forms new condensing droplets due to the falling of large droplets under gravity effect,

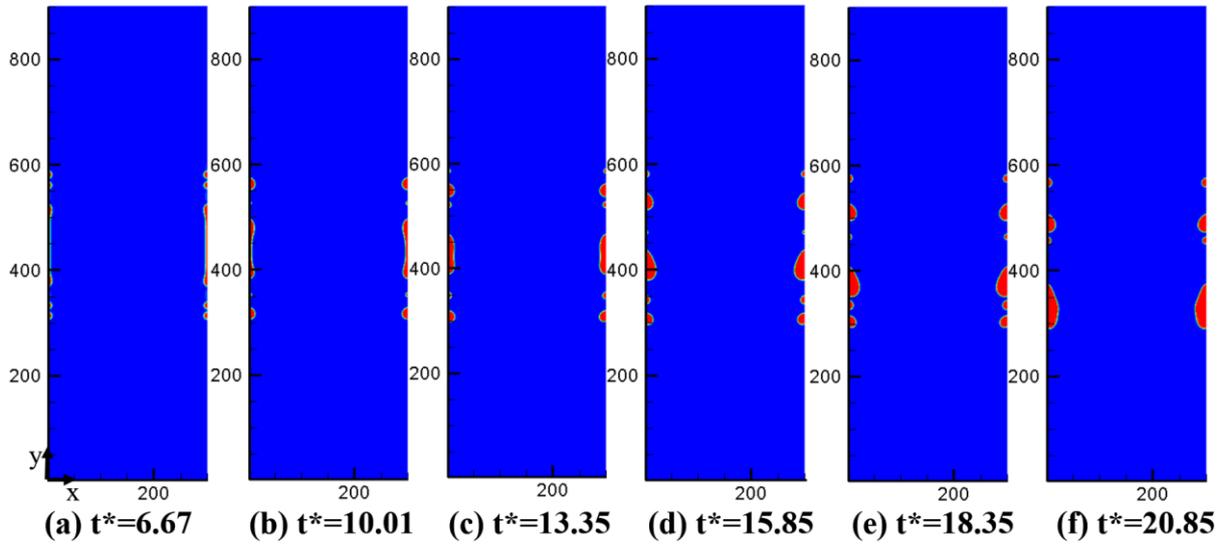

(a) t*=6.67  (b) t*=10.01  (c) t*=13.35  (d) t*=15.85  (e) t*=18.35  (f) t*=20.85

Fig. 7. Two-phase distribution of saturated steam condensation after further increasing the wall contact angle
(Case #2: $Ja = 0.244$, $\theta_w = 130°$)

Next, we change the wetting characteristics of the surface to a super-hydrophobic surface and set the triple-phase contact line as $\theta_w = 155°$, but the degree of wall subcooling still remains $Ja = 0.244$. Fig.8 demonstrates the dynamics behaviors of saturated vapor condensation under a super-hydrophobic subcooling surface. Fig.8(a-c) also gives two-phase contours in the same dimensionless time corresponding to Fig.6(a-c) and Fig.7(a-c), and from the figure we can clearly observe that at t*=6.67, due to the increase of wetting characteristics of the surface, a lot of fines droplets grow on the subcooling surface, it is definitely different from the Fig.6(a) and Fig.7(a) at the same dimensionless time. Immediately afterwards, the liquid film in Fig. 8(b) quickly converges and forms large droplets (t*=10.01, Y=425). As presented in Fig.8(c), the tiny droplets between Y=400-600 grow further on the subcooled wall, while the droplets in the region of Y=200-400 begin to slide down under gravity (t*=15.01), and further merged into one droplet at t*=16.69. Finally, the droplets in the region of Y=400-500 as shown in Fig.8(f) also coalesce into large droplets and fall along the wall. Due to the larger wall contact angle, the head of the falling droplets is significantly away from the wall and forms obviously contact angle hysteresis.

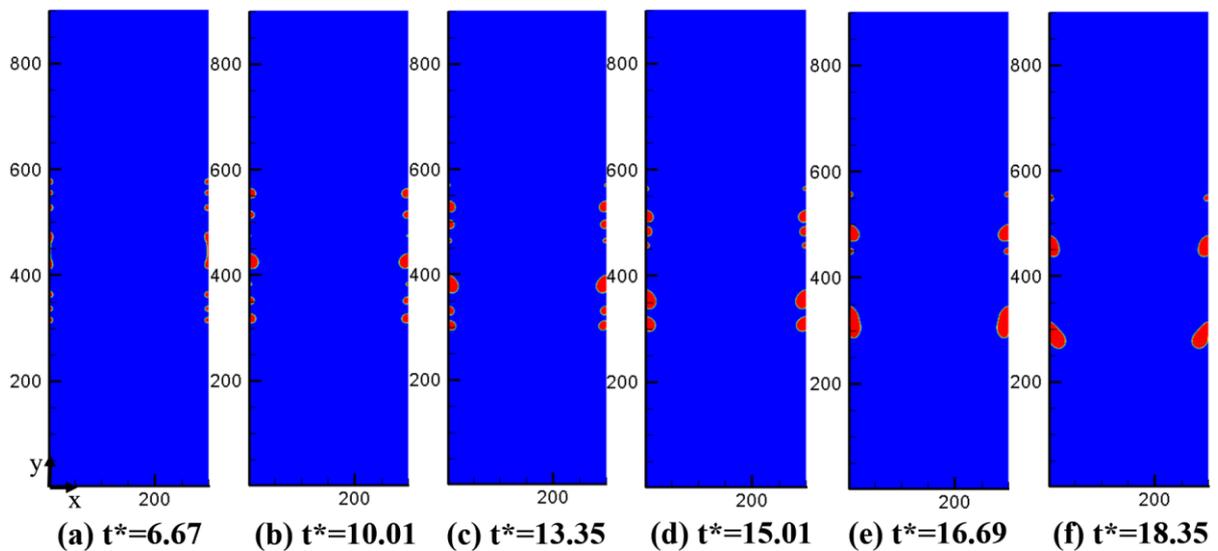

(a) t*=6.67  (b) t*=10.01  (c) t*=13.35  (d) t*=15.01  (e) t*=16.69  (f) t*=18.35

*4.2.2. Velocity vector of computational domain*

To further understand the condensation mechanism of saturated steam under the subcooled wall, Fig. 9 presents the transient velocity vector of the saturated steam during the condensation at the same moment corresponding to Fig.7, and the blue color marked in the figure indicates the interface of the condensate droplets. As can be seen from the figure, at t*=6.67, with the formation of condensation droplets on the wall, the ambient gas flows toward the thin liquid film and droplets. In the region of red dotted line when t*=10.01, saturated vapor flow can be clearly seen at the interface of the droplets, and it can be found that the vapor mainly converges near the triple-phase line while the vapor above the liquid film has almost no flow. Immediately afterwards, it can be clearly seen the vortex flow in the red region in the Fig. 9(c), which is caused by the droplet movement and the effect of viscous. The vapor around the droplet presented by Fig.9(d) continues to converge near the interface of the droplet, and a vortex is formed in the red marking area, and as the droplet further grows, the droplets are coalesced together (t*=18.35). The volume of droplet gets bigger and bigger and falls along the wall under the effect of gravity. Due to the falling of the droplet, the vapor in the red marked area shown in Fig.9(f) forms a larger vortex, which is consistent with the vector contour of the drop falling along the wall.

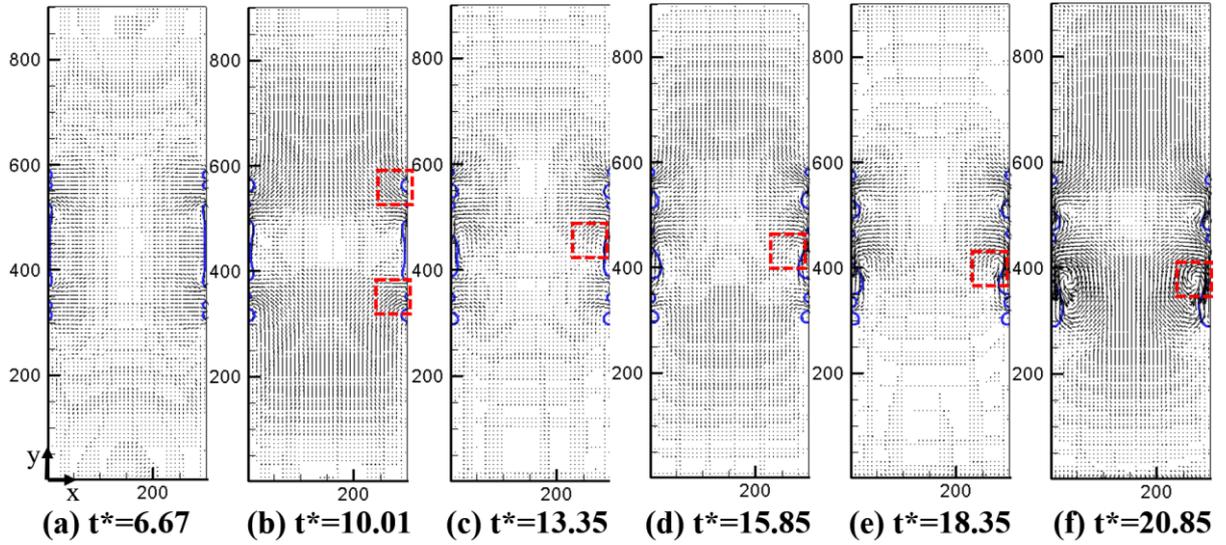

(a) t*=6.67  (b) t*=10.01  (c) t*=13.35  (d) t*=15.85  (e) t*=18.35  (f) t*=20.85

*Fig. 9. Velocity field of the vapor condensation (Case 3#: Ja =0.244, $\theta_w$ =155°)*

*4.2.3. Temperature field and local heat flux*

In this part, we further comprehend the temperature field and transient heat flux of the saturated steam condensation on the homogeneous subcooled wall. Following the Ref. [14], the local heat flux is dimensioned by the following equation:

$$q*(x,y,t) = \frac{(-\lambda \frac{\partial T}{\partial y})}{q_0} = \frac{1}{\rho_l \upsilon_l h_{fg}} \sqrt{\frac{\sigma}{g(\rho_l - \rho_g)}} (-\lambda \frac{\partial T}{\partial y})\bigg|_{x=0} \quad (34)$$

where $q_0 = \mu_l h_{fg} \sqrt{g(\rho_l - \rho_g)/\sigma}$ is the reference condensation heat flux and $q_0^{lu}$ was assumed as 0.01808 in this paper.

Fig. 10 demonstrates the non-dimensional temperature field of the super-hydrophobic surface condensation corresponding to Fig.8(a), (b), (c) and (f), where the isotherm is shown at the same time, and the white line indicates the droplet interface. From the figure, we can observe that at t*=6.67, the temperature above the liquid film in the middle of the condensing region is significantly higher than that in other regions, because the spread area of the liquid film is larger at this time, a thermal resistance forms a thermal resistance and which suppresses the further heat transfer of subcooled wall cooling, while the temperature above the droplets at both ends of the condensing region is consistent. As the cooling of the subcooling wall further released during the condensation, the temperature in the middle area is slowly decreased (t*=10.01). From the Fig.10(c), it is found that the isothermal lines in the droplets have many crowns, and one can observe that the temperature isotherms are more crowded in the triple-phase lines than in another region where no phase change forms. Finally, with the falling of the droplets as shown in Fig.10(f), the cooling is transferred to the gas through the gas region on the subcooled surface, and there is a high temperature above the droplet, so it can be concluded that the droplet indirectly hinders the heat transfer of the cooling to the interior.

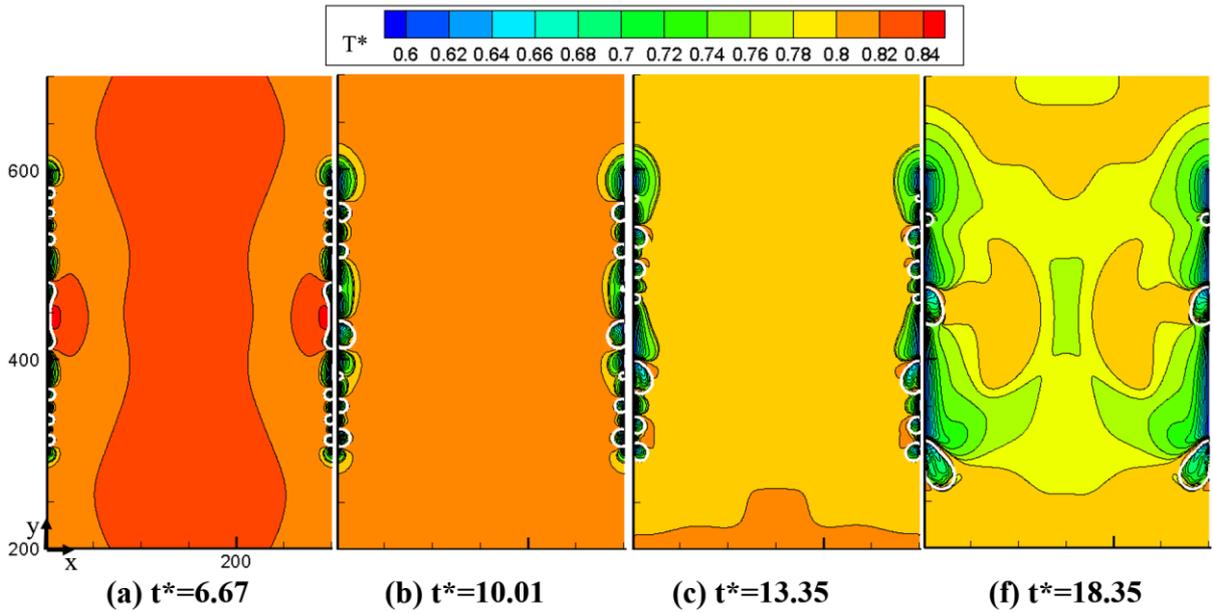

Fig. 10. The dimensionless temperature field ($T^* = T/T_{cr}$) changes with time in the calculation domain (0< $Nx$ <300, 200< $Ny$ <700), it should be noted that $\theta_w$ =155°, $Ja$ =0.24

In the meanwhile, Fig. 11(a) and (b) illustrates the local heat flux distributions corresponding to Fig. 6(a) and Fig. 8(a) at dimensionless time t* = 6.67, respectively. As can be seen from Fig.11 (a), there is a high condensation heat flux in the region of small droplet growth, and the heat flux near the triple-phase line region of the droplet is much higher than that in other regions, while the heat flux of in the middle of the subcooled surface remains constant, which is consistent with the conclusion of the experimental results in the Yanadori [49]. Similarly, we observe the same results in Fig. 11(b), but due to the smaller droplets growth on the super-hydrophobic surface, there are many local peak heat flux appeared in the subcooled surface, which is due to the formation of triple-phase lines on the hydrophobic wall, that is the reason why isotherms are more crowded there. Meanwhile, in Fig.11 (b), the area of constant heat flux is also reduced due to the decreasing of the liquid film area.

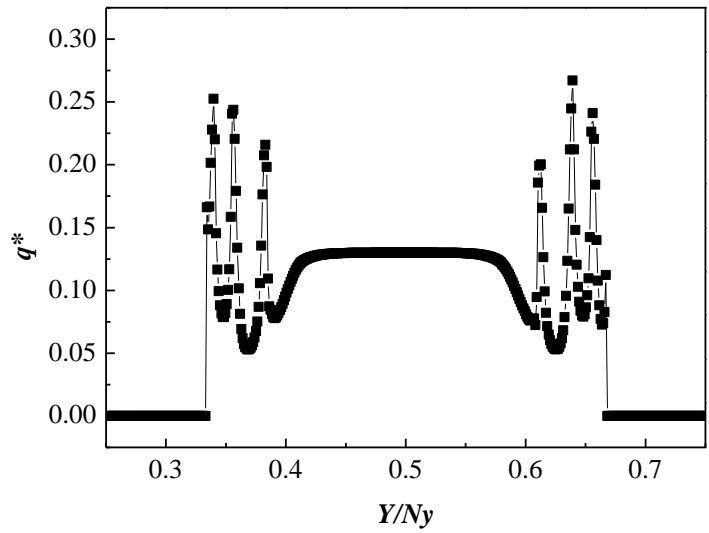

(a) Case #1   $Ja$ =0.244,  $\theta_w$ =115°

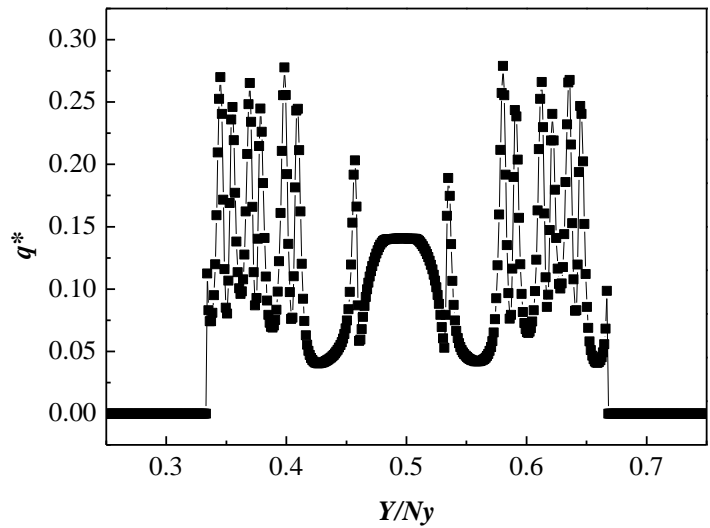

(b) Case #3   $Ja$ =0.244,  $\theta_w$ =155°

Fig. 11. Local heat flux distribution on the homogeneous wall with different wetting characteristics on the subcooled wall at the same time (t*=6.67).

*4.3. Growth, coalescence, falling and heat flux of saturated vapor condensation on heterogeneous hydrophobic-hydrophilic subcooling surface under the gravity*

In this section, we turn our attention to the saturated vapor condensation on the heterogeneous subcooled surface. Effects of wall wetting characteristics, the number of hydrophilic spots and the length of the hydrophilic wall on the saturated vapor condensation will be investigated. In order to clearly discuss the influence of heterogeneous wall structure, Fig. 12 gives the geometric size and the distribution of the hydrophilic spots on the heterogeneous wall, and the yellow areas represent the hydrophilic spots, while the blue areas indicate the hydrophobic surface, and the length of the hydrophilic spots and the hydrophobic wall are $L_s$ and $L_w$, respectively.

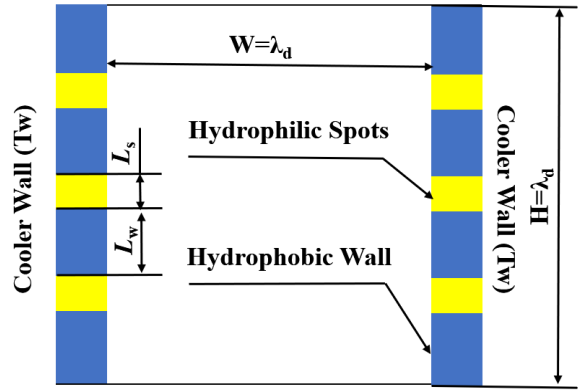

*Fig. 12. schematic of the structure of heterogonous subcooling wall*

*4.3.1. Droplet's growth and coalescence in the heterogeneous subcooling wall under vertical gravity*

Fig. 13 illustrates the condensation process of the saturated vapor when the number of hydrophilic spots is three and the length of the hydrophilic wall is $L_s$ =0.33, but the contact angle of the hydrophobic wall and the subcooling of the wall remain the same (Case 4#, =0.244, =65, =155). It can be clearly seen from the Fig. 13 that there are many tine droplets growth on the heterogeneous wall at the same time of Fig.7(a), rather than form a thin liquid film in the middle of the subcooled surface, so this can greatly improve the formation of the dropwise condensation. Since the influence of latent heat further released during the vapor condensation, droplets grow up (t*=10.01) and some droplets begin to coalesce at t*=13.35.

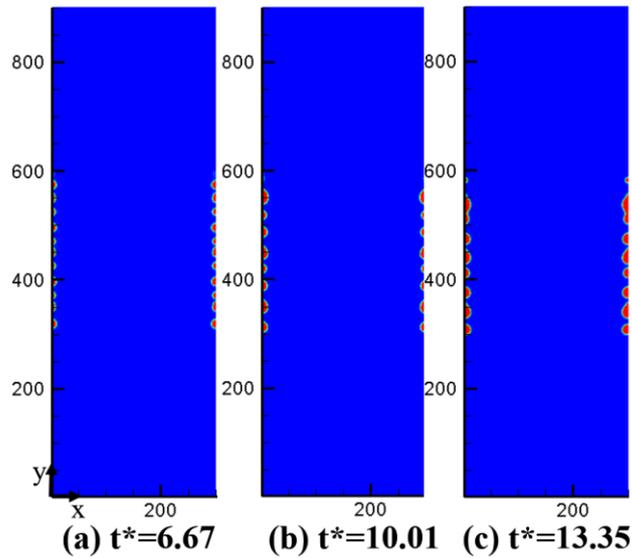

Fig. 13. *Snapshots of droplet's growth, coalescence and falling under the heterogeneous subcooling wall under gravity (Case 4#, $Ja$ =0.244, $\theta_s$ =65°, $\theta_w$ =135°)*

In order to further understand the difference of heat flux between homogeneous and heterogeneous walls during condensation, Fig. 14 demonstrates the local heat flux distribution corresponding to Fig.7 (a) and Fig.13 (a) at the same dimensionless time t*=6.67. One can clearly observe that there are many peaks and valleys on the local heat flux on the heterogeneous wall, while the heat flux is almost constant on the homogeneous wall due to the formation of the filmwise condensation in the middle of the subcooled wall. Therefore, it is proved that at the initial stage of condensation, a better dropwise condensation can be formed on the heterogeneous wall.

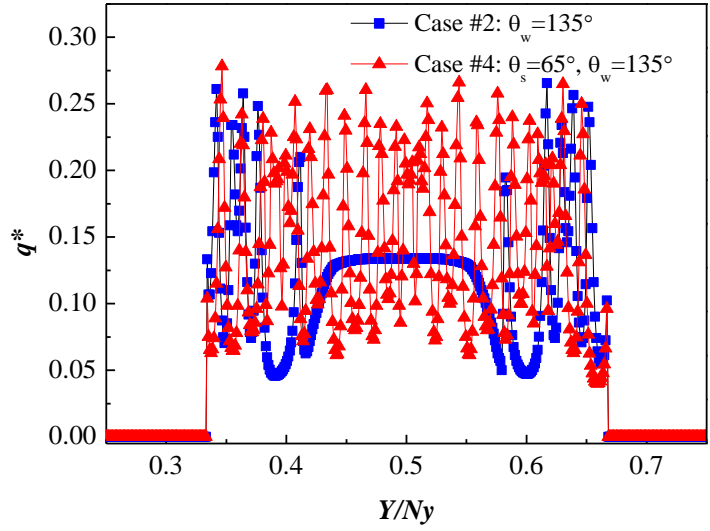

Fig. 14. Local heat flux on the homogenous and heterogeneous wall at t*=6.67 under the same subcooling ( $Ja$ =0.244)

*4.3.2. Effects of Jacob number*

Fig.15 presents the transient heat flux curve of saturated vapor under heterogeneous surface with a wide range of $Ja$ number ( $Ja$ =0.05-0.26) in Case #5, and it is noteworthy that the transient average heat flux is calculated by the following equation:

$$q*_{ave-t} = \frac{1}{\lambda_d} \int_{Ny/2-\lambda_d/2}^{Ny/2+\lambda_d/2} q*(0,y,t)dy \qquad (35)$$

It can clearly observe from the Fig.15 that the condensation process of saturated vapor can be divided into several stages: i) Regime A-B, the natural convection process that occurs when the wall subcooling is low, ii) Regime C-D, as the subcooling of the wall increases, the saturated vapor begins to condense, and as the degree of subcooling increases, the average heat flux increases accordingly, but the increasing slope gradually decreases, iii) Regime D-E, due to the further increase of the degree of subcooling of the wall, resulting in more filmwise condensation on the vertical wall, so the average condensation heat flux decreases at this stage, iv) Regime E-F, in this stage, as the degree of coldness of the cold wall is further increased, the coldness of the wall surface is mainly conducted by means of heat conduction between subcooled wall and liquid film, so that the heat flux density gradually increases. This is consistent with the results obtained in the condensation experiments [49, 50] and numerical results [14].

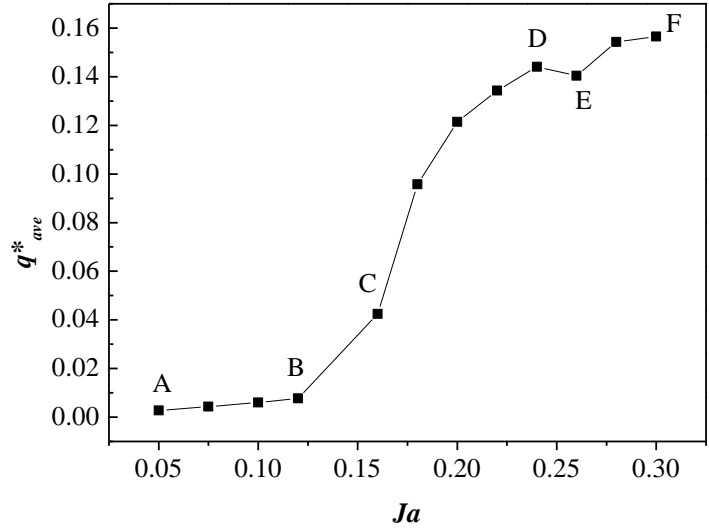

Fig. 15. Transient average heat flux of different subcooled surfaces at t*=6.67 (Case 5#, $\theta_s$ =65°, $\theta_w$ =135°, Ja =0.05-0.26)

In order to further investigate the variation of heat flux during the entire condensation process on heterogeneous vertical subcooled wall under different wetting characteristics. Fig. 16 performs the condensation heat flux curve under different subcooled surfaces, where the average heat flux during the entire condensation process was calculated according to the Ref. [14], and the equation is given as follows:

$$q^*_{ave-t} = \frac{1}{\lambda_d (t_b - t_a)} \int_{t_a}^{t_b} \int_{Ny/2-\lambda_d/2}^{Ny/2+\lambda_d/2} q^*(1, y, t) dy dt \qquad (36)$$

Fig.16 presents the average condensation heat flux on different subcooled heterogeneous surfaces when the number of hydrophilic spots is three and the contact angle of hydrophilic surface keep constant ($\theta_s$=65°). As can be seen from the figure, since the natural convection heat transfer occurs only in the A-B stage where the degree of subcooling is small, the wetting characteristics of the hydrophobic surface have slightly effect on the average heat flux, therefore the two lines almost coincide. However, as the degree of the wall subcooling increases, the condensate droplets begin to appear on the wall, and from the Fig.16, it is clear concluded that the average condensation heat flux increases with the increase of hydrophobic surface wettability, which is in agreement with the numerical simulation results in the Ref. [14]. Especially at Ja=0.16, the non-dimensional average heat flux of Case #5 is 0.0634, while the dimensionless average heat flux of Case #6 is only 0.0420, and the difference reachs 0.0214, but the gap of dimensionless average heat flux decreases with the increase the subcooling, even at Ja=0.26, the average heat flux of case #6 exceeds Case #5. This is because when the wall subcooling increases, the surface with poor wettability is more conducive to the occurrence of dropwise condensation. At the same time, under the gravitational force, the adhesion of the wall to droplets becomes smaller, which is more conducive to the droplet falling along wall, thus improving the further condensation of the saturated vapor. This conclusion is also consistent with the literature [14].

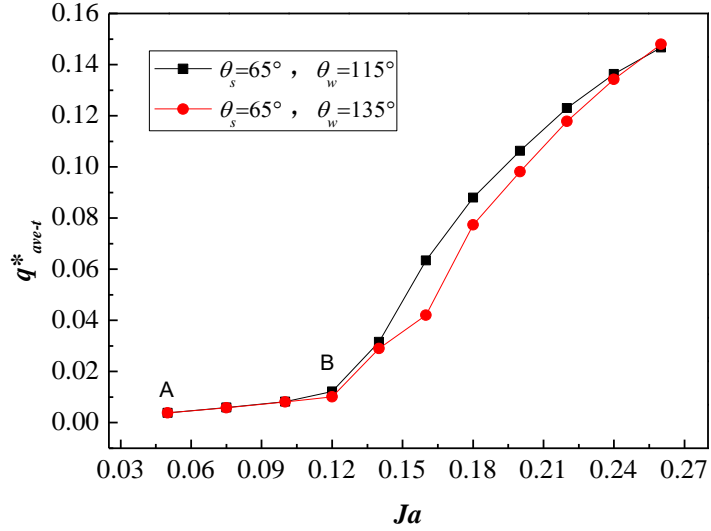

Fig. 16. Effects of wetting ability of hydrophobic wall on the vapor condensation of heterogeneous wall with a wide range of subcooling (Case 5#, $\theta_s=65°$, $\theta_w=115°$ and Case 6#, $\theta_s=65°$, $\theta_w=135°$)

*4.3.3. Effects of the length and number spots of hydrophilic wall*

In this section, we further discuss the effect of the length and the number spots of the hydrophilic surface on the condensation of saturated vapor, but the contact angle of the hydrophilic and hydrophobic walls remain at 65° and 135°, respectively. Fig.17 presents the variation of the average heat flux with $Ja$ number under the condition that the length of the hydrophilic wall is $L_s=0.33$, $L_s=0.66$, $L_s=0.99$, respectively, which corresponds to Case #6, #8 and 9#. From the figure, we can clearly found that when the surface $Ja<0.1$, the average heat flux under the three structures is kept constant, because the convection heat transfer occurs under the low subcooling. As the degree of subcooling increases, when $Ja=0.12$, Case #6 still only undergoes convection heat transfer, while the average heat flux of Case #8 and #9 has suddenly increased. This indicates that at this moment, condensation has already occurred at Case #8 and #9. Therefore, it can be concluded that the vapor condensation will be easily formation with increase in length of the hydrophilic wall. As the degree of subcooling surface increases, condensation occurs in all three structures, but when $Ja$ number remains the same, the structure with larger hydrophilic wall length has a larger average heat flux density, however, as the $Ja$ number further increases, the gap of average heat flux is slowly decreasing. This is because, with the increase in the number of hydrophilic wall surface, although it is beneficial to the growth of the condensate droplet, the increasing effect is not the main factor affecting the heat flux, and the excessive adhesion of the hydrophilic wall to the droplet becomes stronger, which is not conducive to the falling of condensate droplets, but becomes a barrier to further condensation of saturated vapor.

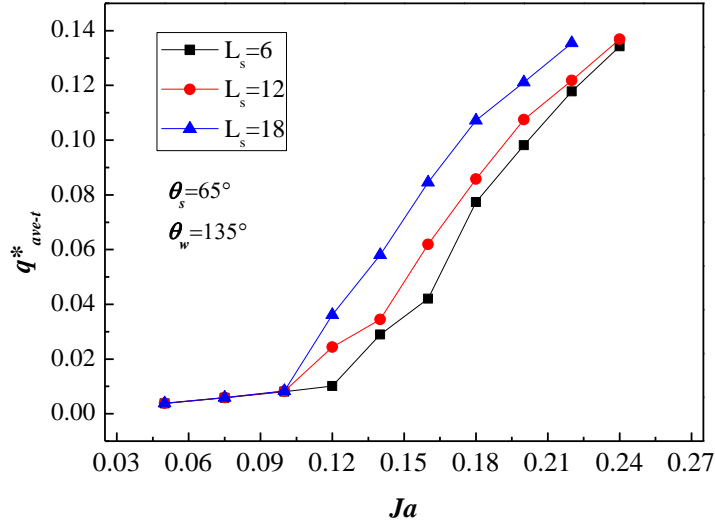

Fig. 17. Effects of the length of hydrophilic surface on the condensation cure

The relationship between the number of hydrophilic spots and the average heat flux is given by Fig. 18. From the figure we can observe that at $Ja<0.12$, the two cures coincide because no condensation heat transfer occurs. When the subcooling is in the range of $0.12<Ja<0.18$, on can observed that the average heat flux of the wall structure with more hydrophilic spots is higher, but as the degree of subcooling surface further increases, the effects of the number of hydrophilic spots is gradually decreasing, which causes the coincidence of the heat flux curves at lager subcooling wall.

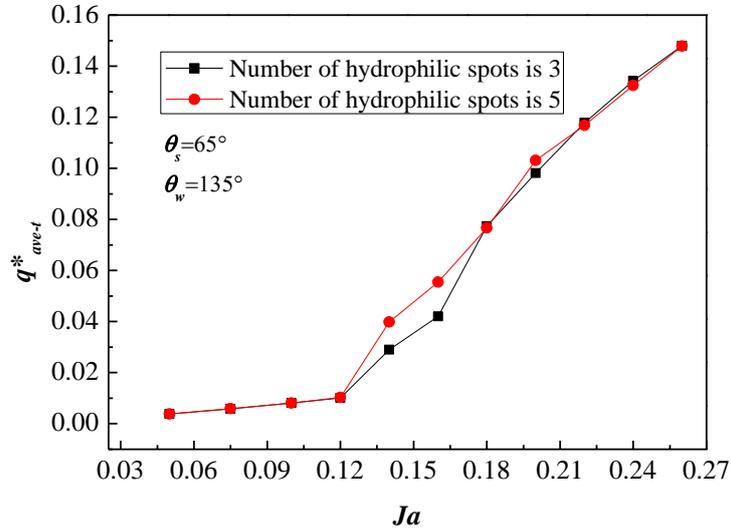

Fig. 18. Effect of the number of hydrophilic on the average heat flux of the vapor condensation with $\theta_s=65°$, $\theta_w=135°$

## 5. Conclusions

In this paper, the improved thermal pseudopotential MRT-LM phase-change model coupling FDM to solve phase change was developed to perform numerical simulation on the vapor condensation on the vertical subcooled wall with gravity. Firstly, the phase-change model was verified via the droplet evaporation and the wetting behaviors on the flat wall, which proves the accuracy and reliability of this model. Subsequently the model was employed to simulate the saturated vapor condensation on the homogeneous and heterogeneous walls under gravity, the growth, coalescence and falling process of the condensate droplets, and the temperature field, the local heat flux and the average heat flux and the structure of heterogeneous surface also are investigated. The following conclusions can be obtained from this paper:

a) For the saturated vapor condensation on the homogeneous subcooled wall, it is easier for dropwise condensation formation with decreasing the wettability of the subcooled surface. and the droplets are easier to fall due to the smaller adhesion of the wall to droplet under the gravity.

b) It can be seen from the velocity vector formed during the condensation that the saturated vapor mainly converges near the triple-phase line and grows into larger droplets. When the droplets further fall, the gas around the droplets will form a vortex due to the effect of viscous force.

c) For the formation of more fine droplets in the condensation process, the local heat flux has more peak heat flux, and at the same time there is a higher condensation heat flux near the triple-phase contact line of the droplet than in other regions.

d) The heterogeneous cooler wall is more conductive to form dropwise condensation at the initial stage of condensation than the homogeneous wall surface, and can further generate and coalesce small droplets at the hydrophilic spots.

e) As the degree of wall subcooling increases, natural convection will occur firstly, and then the condensation process occurs. However, when the degree of wall subcooling further increases, there are three stages for the transient average heat flux: it increases first, then decreases and finally increase with the increase of wall subcooling.

f) When the wall subcooling is small, increasing the hydrophilic length of the heterogeneous wall will be conducive to vapor condensation, and when the wall has a high wall subcooling, which causes the droplets adhere wall more easily and it is not conducive to the falling of the droplets.

Since this paper only simulates the saturated vapor condensation in the two-dimensional subcooled wall, the vapor condensation in three-dimensional domain will be further investigated in the next work.

**Acknowledge**

This work was supported by National Natural Science Foundation of China (No. 11562011, No. 51566012), Graduate Innovation Special Foundation of Jiangxi Province (No. YC2017-S056) and Jiangxi Provincial Department of Science and Technology (No. 2009BGA01800). The authors are also grateful for the support from the New Faculty Support program at The University of Texas Rio Grande Valley. The first author is grateful to the China Scholarship Council (CSC) for financially support (No.201806820023). Special thanks are also due to Prof. Qing. Li from Central South University for his great help and valuable discussions on the theory and programing of this model.